\documentclass[%
 aip,
 amsmath,amssymb,
 reprint,%
]{revtex4-1}
\usepackage{graphicx}
\usepackage{dcolumn}
\usepackage{bm}
\usepackage{xcolor}
\usepackage{comment}

\usepackage[utf8]{inputenc}
\usepackage[T1]{fontenc}
\usepackage{mathptmx}
\usepackage{etoolbox}
\makeatletter
\def\@email#1#2{%
 \endgroup
 \patchcmd{\titleblock@produce}
  {\frontmatter@RRAPformat}
  {\frontmatter@RRAPformat{\produce@RRAP{*#1\href{mailto:#2}{#2}}}\frontmatter@RRAPformat}
  {}{}
}%
\makeatother
\begin{document}

\preprint{AIP/123-QED}

\title[Multiplexing-based control of stochastic resonance]{Multiplexing-based control of stochastic resonance}
\author{Vladimir V. Semenov}
\email{semenov.v.v.ssu@gmail.com}
\affiliation{Technische Universit\"{a}t Berlin, 10623 Berlin, Germany}
\affiliation{Saratov State University, 410012 Saratov, Russia}

\author{Anna Zakharova}
\affiliation{Technische Universit\"{a}t Berlin, 10623 Berlin, Germany}

\date{\today}

\begin{abstract}
We show that multiplexing\footnote{Here, the term 'multiplexing' means a special network topology where a one-layer network is connected to another one-layer networks through coupling between replica nodes. In the present paper, this term does not refers to the  signal processing issues and telecommunications.} allows to control noise-induced dynamics of multilayer networks in the regime of stochastic resonance. We illustrate this effect on an example of two- and multi-layer networks of bistable overdamped oscillators. In particular, we demonstrate that multiplexing suppresses the effect of stochastic resonance, if the periodic forcing is present in only one layer. In contrast, the multiplexing allows to enhance the stochastic resonance, if the periodic forcing and noise are present in all the interacting layers. In such a case the impact of multiplexing has a resonant character: the most pronounced effect of stochastic resonance is achieved for an appropriate intermediate value of coupling strength between the layers. Moreover, multiplexing-induced enhancement of the stochastic resonance can become more pronounced for increasing number of coupled layers. To visualize the revealed phenomena, we use the evolution of the dependence of the signal-to-noise ratio on the noise intensity for varying strength of coupling between the layers. 
\end{abstract}

\maketitle

\begin{quotation}
Stochastic resonance is common in non-autonomous bistable dynamical systems and represents an interdisciplinary noise-induced phenomenon. This effect is manifested by the increasing regularity of the stochastic dynamical system response to an input signal for increasing noise intensity. Since the effect of stochastic resonance is widely used in practice (for instance, see Refs. \cite{zheng2014,shuyao2016,qiao2019,roy2019} and the list of references in the review by L. Gammaitoni\cite{gammaitoni1998}), the issues addressing its control are relevant in the context of both applied and theoretical science. Here, we propose a mechanism of controlling stochastic resonance in multilayer networks of coupled bistable oscillators. We show that connecting a one-layer network to another one-layer network through coupling between replica nodes, i.e., multiplexing, provides a tool for controlling stochastic resonance. In particular, we demonstrate that stochastic resonance can be more or less pronounced depending on the coupling between the layers and the properties of the periodic forcing impact. We expect a broad variety of potential applications of our results, since the issues of stochastic multilayer network dynamics are actively developing, especially in the context of artificial intelligence.
\end{quotation}

\section{Introduction}
\label{intro}
The effect of stochastic resonance \cite{gammaitoni1998,anishchenko1999} is observed in a broad variety of non-autonomous dynamical systems and unites a wide spectrum of stochastic processes observed in the frameworks of laser physics \cite{mcnamara1988,fioretti1993,bartussek1994}, electronics \cite{luchinsky1999-1,luchinsky1999-2,calvo2006}, chemistry \cite{guderian1996,foerster1996,hohmann1996,davtyan2016,leonard1994,yang1999}, biology \cite{haenggi2002,hong2006,mcdonnell2009}, geophysics \cite{stone1998}, climatology \cite{benzi1982,nicolis1981,nicolis1982}, population dynamics \cite{valenti2004,park2021,mubayi2019}, economics \cite{traulsen2004}. The issues addressing stochastic resonance are not limited by the single-system stochastic dynamics, but also involve the collective oscillatory behaviour \cite{gosak2011,tang2012}, effects in spatially-extended systems \cite{benzi1985,jung1995,wio1996,wio2002,lai2009} and time-delay oscillators considered as a spatially-extended system by means of virtual space-time representation \cite{semenov2016}. This phenomenon can be observed in itself as well as be accompanied by effects of pattern formation such as spiral wave excitation \cite{jung1995} and chimera states \cite{semenov2016}. 

There are known approaches for controlling the characteristics of noise-induced oscillations in the regime of stochastic resonance: introducing time-delayed feedback \cite{mei2009,jia2010}, tuning the statistical characteristics of noise such as the correlation time \cite{haenggi1993}, using additive and multiplicative noise sources simultaneously present in a dynamical system \cite{qiao2016}, varying the coupling strength in the case of coupled oscillators \cite{neiman1995,nicolis2017}. 

Recently, control schemes based on multiplexing have been reported for coupled deterministic dynamical systems with static inter-layer topology. In particular, multiplexing has been shown to regulate the dynamics in the regime of chimera state \cite{ZAK20} or solitary state \cite{SCH21} and to influence the interplay between these two states of the network \cite{MIK19,RYB21}. Additionally, topological asymmetries in multilayer networks have been found to induce regularity, i.e., rescue a nonlinear system from chaotic dynamics by establishing stable periodic orbits and equilibria (so-called asymmetry-induced order) \cite{MED21}. The dynamic inter-layer topology as well can have a significant impact, e.g., on the interlayer synchronization: random switching of the inter-layer links improves the system synchronizability allowing the layers to synchronize at lower inter-layer connectivity \cite{ESE21}. 

Moreover, noisy modulation of the inter-layer coupling strength, called multiplexing noise, has been shown to control the inter-layer synchronization of spatio--temporal patterns in multilayer networks \cite{VAD20,RYB22}. However, the phenomenon of stochastic resonance and, in particular, its control in multilayer networks has not been yet investigated.

In the present work, we propose a new control method which allows to enhance or suppress stochastic resonance and can be realized in multilayer networks. The discussed mechanism consists in the adjustment of the coupling between the layers. A similar approach \cite{semenova2018,MAS21} has been successfully applied to control the effect of coherence resonance 
\cite{PIK97,MAS17} observed in a multilayer network of excitable oscillators including control of coherence resonance by self-induced stochastic resonance \cite{yamakou2019}. Moreover, as demonstrated in Ref.\cite{yamakou2022}, the multiplexing technique provides for controlling self-induced stochastic resonance and coherence resonance even when multiplexing connections are adaptive. Thus, in the present article, our goal is on one hand to develop a tool to control stochastic resonance. On the other hand we aim to generalize the role of multiplexing for stochastic resonant phenomena which consists in providing a mechanism for enhancement and suppression of the noise-induced regularity.

Nowadays, the issues addressing stochastic multilayer network dynamics are actively studied in the context of deep learning \cite{semenova2019,semenova2022,semenova2022-2}. Since the property of bistability is easily achieved in various kinds of artificial neural networks (for instance, see Refs. \cite{stern2014,vecoven2021} where bistable neural networks with tanh-nonlinearity are considered), one can expect the occurrence of stochastic resonance in such systems, which can potentially affect neural network characteristics and performance. Due to this fact, we expect that the presented results would be interesting for experts in artificial intelligence besides specialists in theory of stochastic processes and nonlinear dynamics. In addition, we hope the presented results will be useful for researchers of spin networks where the bistability is manifested as the coexistence of symmetric spin-up and spin-down states.

\section{Single-layer dynamics}
\begin{figure}[t]
\centering
\includegraphics[width=0.48\textwidth]{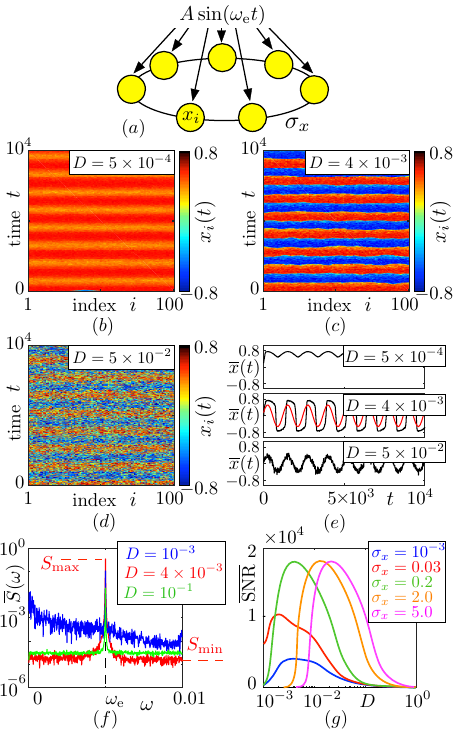} 
\caption{Stochastic resonance in a single-layer network Eqs. (\ref{single-layer}). (a) Schematic representation of a one-layer network (layer $x$). (b)-(d) Spatio-temporal dynamics for increasing noise intensity $D$; (e) Time realization of the dynamical variable ensemble average $\overline{x}(t)$ for varying noise intensity $D$. The red curve displays the external forcing signal $A\sin(\omega_{\text{e}} t)$; (f) Evolution of the power spectrum averaged over the ensemble $\overline{S}(\omega)$ for varying noise intensity $D$. Other parameters: $m=0.25$, $\sigma_x=0.2$, $A=0.04$, $\omega_{\text{e}}=0.005$; (g) Dependencies $\text{SNR}(D)$ for varying $\sigma_x$. Other parameters: the same as in panels (a)-(f).}
\label{fig1}
\end{figure}

Before focusing on the multiplexing impact, we consider a single ring of coupled overdamped oscillators exhibiting the coexistence of two stable steady states  [Fig.~\ref{fig1}~(a)] to allow for further comparative analysis of isolated- and coupled-layer dynamics. The oscillators in the layer represent the Kramers oscillator including an additive source of noise which is a classical example of the stochastic bistable system describing Brownian motion in a double-well potential \cite{hanggi1990,kramers1940,freund2003}. All the oscillators are driven by a common external periodic force. System equations take the following form:
\begin{equation}
\label{single-layer}
\begin{array}{l}
\dfrac{dx_{i}}{dt}=mx_{i}-x_{i}^3+A\sin(\omega_{\text{e}} t)+\sqrt{2D}n_i(t)\\
+\dfrac{\sigma_{x}}{2}\sum\limits^{i+1}_{j=i-1}(x_{j}-x_{i}), \\
\end{array}
\end{equation}
where $x_i$ are the dynamical variables, $i=1, 2, ..., N$ with $N$ being the total number of elements in the layer. In this study, all the network layers consist of $N=100$ elements.  The strength of the coupling within the layer (intra-layer coupling) is given by $\sigma_{x}$. Parameter $m$ determines the dynamics of an individual network element. It defines whether the individual element is monostable ($m<0$) or bistable ($m>0$). In the current study, we assume that all the elements are in the bistable regime. Further, $\sqrt{2D}n_i(t)\in\mathbb{R}$ is Gaussian white noise with intensity $D$, i.e., $<n_i(t)>=0$ and $<n_i(t)n_{j}(t)>=\delta_{ij}\delta(t-t')$, $\forall i,j$. The ensemble under study is considered by means of numerical simulations. Numerical modelling of the model is carried out by integration of studied differential equations using the Heun method \cite{mannella2002} with the time step $\Delta t=0.001$ and the total integration time $t_{\text{total}}=10^6$.

For a single ring of locally coupled bistable oscillators we study the role of noise intensity $D$ and coupling strength $\sigma_{x}$ for fixed parameter $m=0.25$, corresponding to the coexistence of two stable steady states $x^*_{1,2}=\pm0.5$ in the phase space of a single oscillator. The intra-layer coupling strength varies in the range $\sigma_x \in [0.001:5.0]$. The external periodic forcing amplitude is fixed, $A=0.04$, such that the amplitude is less than the threshold value which induces transitions between the states $x^*_{1,2}$ in the ensemble. The external forcing frequency is assumed to be low, which allows to make the stochastic resonance more pronounced, $\omega_{\text{e}}=0.005$. The used initial conditions are chosen to be random and uniformly distributed in the range  $x_i(t=0)\in[0.25:0.75]$. These initial conditions correspond to the basin of attraction of the steady state $x^*_{1}=0.5$. Such combination of the initial conditions allows to avoid the impact of multistability exhibited by the system, which is especially significant for weak intra-layer coupling.

For the chosen set of parameters and initial conditions, the oscillators of system Eqs. (\ref{single-layer}) exhibit coherent oscillations in the basin of attraction of the steady state $x^*_{1}=0.5$. In such a case, transitions between states $x^*_{1,2}=\pm0.5$ are not observed in the presence of weak noise [Fig.~\ref{fig1}~(b)]. However, noise of larger intensity leads to the increasing of the system's response amplitude, while the spatial coherence persists (as illustrated in Fig.~\ref{fig1}~(b,c), the coupled oscillators exhibit almost simultaneous transitions). For an appropriate noise intensity, one can distinguish almost regular collective oscillations involving transitions between two basins of attraction of states $x^*_{1,2}$ [Fig.~\ref{fig1}~(c)]. Increasing noise intensity even further makes the system's response regularity decrease [Fig.~\ref{fig1}~(d)]. To describe the observed noise-induced dynamics in the same way as it is done for classical stochastic resonance, we introduce the averaged value of a dynamical variable in the ensemble, $\overline{x}(t)=\dfrac{1}{N}\sum\limits_{i=1}^N x_i(t)$, characterizing the global instantaneous state of the ensemble. As illustrated in Fig.~\ref{fig1}~(e), the realizations $\overline{x}(t)$ for varying noise intensity $D$ undergo transformations which are typical for stochastic resonance in a single bistable system: the amplitude of the stochastic system response increases, whereas the oscillations are in phase with the external force (see Fig.~\ref{fig1}~(e) middle panel). To emphasize the similarity of the effects observed in the ensemble with the stochastic resonance in single oscillators, we take into consideration the power spectrum averaged over the ensemble: $\overline{S}(\omega)=\dfrac{1}{N}\sum\limits_{i=1}^N S_i(\omega)$, where $S_i(\omega)$ is the power spectrum of the individual element oscillations $x_i(t)$. Then the power spectrum evolution caused by the noise intensity growth fully corresponds to classical stochastic resonance [Fig.~\ref{fig1}~(f)]. First, the height of the spectral peak at the external forcing frequency increases and the peak becomes most pronounced at certain noise intensity. After that, the system's response to the external periodic forcing becomes less and less regular and one observes inverse transformations of the power spectrum. 

To quantitatively describe the noise-induced dynamics of the system, we introduce the signal-to-noise ratio (SNR). Here, the SNR is introduced in terms of radiophysics and electronics. The power spectra of each oscillator in the ensemble include the spectral peak $S_{\text{max}}$ at the frequency of the external forcing, which is also the main peak in the power spectra. The power spectra also have a minimum $S_{\text{min}}$ close to the main spectral peak ($S_{\text{max}}$ and $S_{\text{min}}$ are schematically illustrated in Fig.~\ref{fig1}~(f) on the example of the averaged spectrum for $D=4\times10^{-3}$). Technically, $S_{\text{min}}$ is calculated as a mean value of spectral components $S(\omega)$ in the range [$0.75\omega_{\text{e}}:0.85\omega_{\text{e}}$]$\land$ [$1.15\omega_{\text{e}}:1.25\omega_{\text{e}}$] which excludes the neighbourhood of the main spectral peak at the external forcing frequency $\omega_{\text{e}}$.
One of the most common SNR's definitions is $\text{SNR}=P_{\text{S}}/P_{\text{N}}$, where $P_{\text{S}}$ is the power of the signal and $P_{\text{N}}$ is the noise power. Then the following formula of $\text{SNR}$ describes the regularity of the experimentally acquired harmonic signal: $\text{SNR}=H_{\text{S}}/H_{\text{N}}$, where $H_{\text{S}}$ is the height of the spectral line above the background noise level in the power spectrum, and $H_{\text{N}}$ is the background noise level close to the resonance frequency $\omega_{\text{e}}$. Thus, in terms of power spectra, the formula for the $\text{SNR}$ takes the form $\text{SNR}=(S_{\text{max}}-S_{\text{min}})/S_{\text{min}}$. The consideration of the $\text{SNR}$ as a function of noise intensity $D$ allows to obtain a non-monotonic curve being a signature of stochastic resonance: there exists an appropriate noise intensity level corresponding to the maximal $\text{SNR}$. To analyze the collective dynamics in ensemble Eqs. (\ref{single-layer}), we compute the power spectra for all the oscillators by using time realizations $x_i(t)$ and then calculate the corresponding SNRs. After that, we extract the mean value of the SNR over the ensemble, $\overline{\text{SNR}}$. 

As depicted in Fig.~\ref{fig1}~(g), the increase of intra-layer coupling strength $\sigma_x$ allows to make the effect of stochastic resonance more pronounced, which is manifested by the increase of SNR's peak values. The coupling-induced enhancement of the stochastic resonance tends to saturation for strong interaction of the oscillators. It is seen in Fig.~\ref{fig1}~(g) that varying the coupling strength does not principally affect the stochastic resonance at $\sigma_x>0.2$ except for the horizontal shift towards larger noise intensity values in dependences $\overline{\text{SNR}}(D)$.

Intuitively, the enhancement of stochastic resonance caused by increasing coupling strength can be explained as follows. In the regime of stochastic resonance the oscillators of the ring Eqs. (\ref{single-layer}) behave such that the oscillations at the frequency of external forcing become the dominant component of the dynamics. Thus, all the oscillators are driven by the external forcing $A\sin(\omega_{\text{e}} t)$ and, at the same time, are under the impact of neighboring oscillators which mainly oscillate at the frequency $\omega_{\text{e}}$. As a result, the spectral peak at the frequency $\omega_{\text{e}}$ increases. 

\section{Interaction of forced and free layers}
\begin{figure}[b]
\centering
\includegraphics[width=0.48\textwidth]{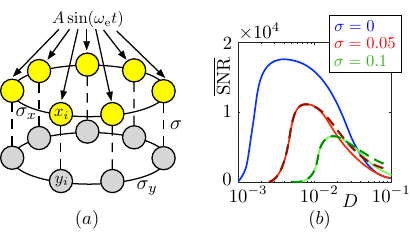} 
\caption{Stochastic resonance in a two-layer network Eqs. (\ref{two-layer-asymmetrical}). (a) Schematic representation of a forced noisy layer $x$ coupled to a free layer $y$; (b) The mean SNR as a function of $D$ in ensemble $x$ (solid curves) and $y$ (dashed curves) for increasing inter-layer coupling strength $\sigma$. Other parameters: $m=0.25$, $\sigma_x=\sigma_y=0.2$, $A=0.04$, $\omega_{\text{e}}=0.005$.}
\label{fig2}
\end{figure}

Next, we consider a two-layer multiplex network depicted in Fig.~\ref{fig2}~(a), where each layer represents a ring of locally coupled bistable oscillators. The oscillators $x_i$ in the first layer are driven by a common external periodic force and contain an additive source of noise, while the second-layer individual oscillators $y_i$ are not under direct action of noise and the periodic force. The system equations take the form
\begin{equation}
\label{two-layer-asymmetrical}
\begin{array}{l}
\dfrac{dx_{i}}{dt}=mx_{i}-x_{i}^3+A\sin(\omega_{\text{e}} t)+\sqrt{2D}n_i(t)\\
+\dfrac{\sigma_{x}}{2}\sum\limits^{i+1}_{j=i-1}(x_{j}-x_{i})+\sigma(y_{i}-x_{i}), \\
\dfrac{dy_{i}}{dt}=my_{i}-y_{i}^3+\dfrac{\sigma_{y}}{2}\sum\limits^{i+1}_{j=i-1}(y_{j}-y_{i})+\sigma(x_{i}-y_{i}). \\
\end{array}
\end{equation}
The strength of the coupling within the layer (intra-layer coupling) is given by $\sigma_{x}$ and $\sigma_{y}$ for the first and second layer, respectively. The coupling between the layers (inter-layer coupling) is bidirectional, diffusive and its strength is characterized by a parameter $\sigma$. We consider a multiplex network, where the layers contain the same number of nodes and the inter-layer links are allowed only for replica nodes, i.e., there is a one-to-one correspondence between the nodes in different layers. 

To demonstrate the action of multiplexing on stochastic resonance, we fix the intra-layer coupling strength $\sigma_{x}=\sigma_{y}=0.2$ and choose all other parameters to be same as in Sec. II. Then we let the inter-layer coupling strength increase starting from zero. The initial conditions are chosen to be similar to those in Sec. II: $x_i(t=0)\in[0.25:0.75]$ and $y_i(t=0)\in[0.25:0.75]$. The obtained results are depicted in Fig.~\ref{fig2}~(b), where the functions $\overline{\text{SNR}}(D)$ are presented for increasing $\sigma$. Similarly to Sec. II, here  $\overline{\text{SNR}}$ is the mean signal-to-noise ratio in a given layer. Since there is no interaction between the layers at $\sigma=0$, for this case Fig.~\ref{fig2}~(b) displays the curve $\overline{\text{SNR}}(D)$ only for the layer $x$. At non-zero $\sigma$, the curves $\overline{\text{SNR}}(D)$ are illustrated for both the first layer $x$ (solid curves) and the second layer $y$ (dashed lines). 
Interestingly, we observe multiplexing-induced stochastic resonance in layer $y$, which receives periodic action and noise only through multiplexing. Moreover, by tuning the inter-layer coupling strength, one can make the effect of stochastic resonance in both layers less and less pronounced by increasing the inter-layer coupling (Fig.~\ref{fig2}~(b)). 

The obtained result can be explained by the fact that multiplexing provides for additional interaction between free and forced bistable elements. The oscillations of the forced elements are characterized by the main spectral peak at the frequency of the periodic forcing and affect the free layer oscillators through the multiplexing. As a result, the SNRs in the layer $y$ increase and tend to the first layer SNRs for increasing inter-layer coupling. At the same time, layer $y$ which has no direct periodic forcing, reduces the SNRs in the layer $x$ with increasing multiplexing strength. 

The multiplexing-induced dynamics we report here (Fig.~\ref{fig2}~(b)) corresponds to the interaction of a stochastic layer with an initially deterministic layer (see Eqs. (\ref{two-layer-asymmetrical})). However, the same results are observed when additive noise is present in both layers. Therefore, one can conclude that the main factor for the observation of multiplexing-based suppression of stochastic resonance is the presence of periodic forcing in either layer while the presence of noise in either or in both layers is not crucial. 

\section{Interacting forced layers}
Further, we study multiplexing networks which consist of identical layers of stochastic oscillators under a common periodic forcing $A\sin(\omega_{\text{e}}t)$. Starting from a two-layer network, we further highlight the impact of a number of interacting layers on the multiplexing-based control of stochastic resonance. Since the interacting layers are identical, the action of multiplexing is illustrated by the evolution of the curve $\overline{\text{SNR}}(D)$ in the first layer $x$ in a two-layer network and in the layer $x_1$ in a multilayer network consisting of $L>2$ layers. The role of multistability is excluded from the consideration. For this reason, we use random initial conditions for all the oscillators uniformly distributed in the range $[0.25:0.75]$, i.e., in one potential well.
\subsection{Two-layer network}
\begin{figure}[t]
\centering
\includegraphics[width=0.48\textwidth]{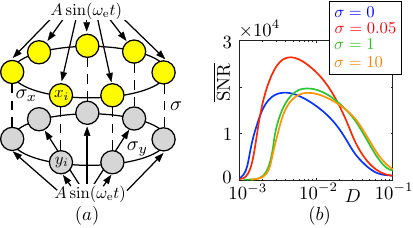} 
\caption{Stochastic resonance in a two-layer network Eqs. (\ref{two-layer-symmetrical}). (a) Schematic representation of two multiplexed noisy forced layers; (b) The mean SNR in ensemble $x$ as a function of $D$ for increasing inter-layer coupling $\sigma$. Other parameters: $m=0.25$, $\sigma_x=\sigma_y=0.2$, $A=0.04$, $\omega_{\text{e}}=0.005$.}
\label{fig3}
\end{figure}

Oscillators of the two-layer network shown in Fig.~\ref{fig3}~(a) are identical and forced both by a common periodic forcing and a source of additive Gaussian white noise. The corresponding model equations under study are
\begin{equation}
\label{two-layer-symmetrical}
\begin{array}{l}
\dfrac{dx_{i}}{dt}=mx_{i}-x_{i}^3+A\sin(\omega_{\text{e}} t)+\sqrt{2D}n_{xi}(t)\\
+\dfrac{\sigma_{x}}{2}\sum\limits^{i+1}_{j=i-1}(x_{j}-x_{i})+\sigma(y_{i}-x_{i}), \\
\dfrac{dy_{i}}{dt}=my_{i}-y_{i}^3+A\sin(\omega_{\text{e}} t)+\sqrt{2D}n_{yi}(t)\\
+\dfrac{\sigma_{y}}{2}\sum\limits^{i+1}_{j=i-1}(y_{j}-y_{i})+\sigma(x_{i}-y_{i}). \\
\end{array}
\end{equation}
Similarly to the Sec. III, we fix the intra-layer interaction strength, $\sigma_{x}=\sigma_{y}=0.2$ and choose all the other parameters correspondingly. Then we again let the inter-layer coupling strength increase starting from zero. The obtained results are illustrated in Fig.~\ref{fig3}~(b) via the evolution of the dependence of the mean signal-to-noise ratio in ensemble $x$ on the noise intensity $\overline{\text{SNR}}(D)$. As demonstrated in Fig.~\ref{fig3}~(b) the increase of the parameter $\sigma$ allows to enhance the effect of stochastic resonance and achieve higher signal-to-noise ratios in comparison with isolated layer dynamics (compare blue and red curves in Fig.~\ref{fig3}~(b)). However, further increasing inter-layer coupling returns the stochastic resonance manifestation to the initial, isolated-layer form (except of horizontal shift, see the blue, green and orange curves in Fig.~\ref{fig3}~(b)). This indicates the resonant character of the multiplexing impact on the noise-induced dynamics: there is an appropriate inter-layer coupling strength corresponding to the most pronounced effect of stochastic resonance. The possible explanation for the resonant inter-layer coupling impact is the redistribution of the interaction strength which occurs due to multiplexing and acts in a similar way as the increase of the intra-layer coupling in the single-layer case. Additional similarity of one- and two-layer case is the saturation occurring for large enough strength of the interaction.

\subsection{Multilayer network with $L>2$ layers}
\begin{figure}[t]
\centering
\includegraphics[width=0.48\textwidth]{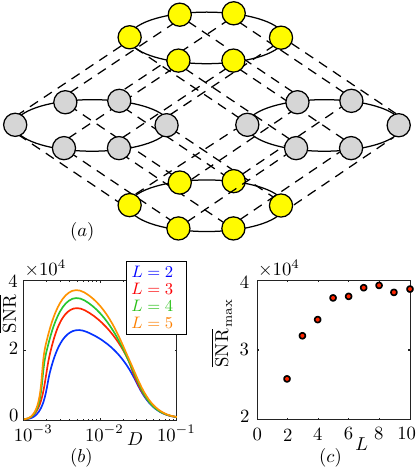} 
\caption{Stochastic resonance in a multilayer network with $L>2$ layers. (a) Schematic representation of a multilayer network with $L=4$ multiplexed layers Eqs. (\ref{multiple-layer-symmetrical}); (b) Curves $\overline{\text{SNR}}(D)$ for fixed parameters $m=0.25$, $\sigma_x=0.2$, $\sigma=0.05$, $A=0.04$, $\omega_{\text{e}}=0.005$ and varying the number of interacting layers; (c) Dependence of the maximal achieved value in the curves $\overline{\text{SNR}}(D)$ on the number of interacting layers. Parameters: the same as for panel (b).}
\label{fig4}
\end{figure}

Suppose that the network under study consists of $L$ identical interacting layers under a common periodic force. The network topology is assumed to be homogeneous, and for this reason the interacting layers form a circle structure as illustrated in Fig.~\ref{fig4}~(a) for four layers. In such a case the system equations take the form written below for $i$-th oscillator in  $l$-th layer:
 \begin{equation}
\label{multiple-layer-symmetrical}
\begin{array}{l}
\dfrac{dx_{i,l}}{dt}=mx_{i,l}-x_{i,l}^3+A\sin(\omega_{\text{e}} t)+\sqrt{2D}n_{i,l}(t)\\
+\dfrac{\sigma_{x}}{2}\sum\limits^{i+1}_{j=i-1}(x_{j,l}-x_{i,l})
+\dfrac{\sigma}{2}\sum\limits^{l+1}_{j=l-1}(x_{i,j}-x_{i,l}).
\end{array}
\end{equation}
All the oscillator parameters and intra-layer coupling strength are the same as in Sec. IV. The inter-layer coupling strength is fixed, $\sigma=0.05$, which corresponds to the most pronounced stochastic resonance in two-layer network Eqs. (\ref{two-layer-symmetrical}) (see the red curve $\overline{\text{SNR}}(D)$ in Fig.~\ref{fig3}~(b)). Then the number of interacting layers is increased starting from two layers. As demonstrated in Fig. ~\ref{fig4}~(b), increasing the number of interacting layers $L$ strengthens the constructive role of multiplexing manifested by the enhancement of stochastic resonance. However, the enhancement of stochastic resonance tends to saturation with increasing $L$. To visualize this fact, we demonstrate the dependence of the maximal achieved value $\overline{\text{SNR}}(D)$ for fixed $L$ and on the number of interacting layers, $L$. As seen from the Fig.~\ref{fig4}~(c), this curve tends to the saturation for $L>5$. The obtained results indicate that one can enhance stochastic resonance in multiplex networks by varying the number of interacting layers. 

The effect of saturation can be explained by the fact that the multilayer network is transformed into a grid structure creating a torus-like surface with increasing number of layers. Then the phenomena discussed above are reduced to stochastic resonance in two-dimensional grids. Thus, it appears feasible that the effect does not fundamentally change with increasing system scale.

\section{Conclusions}
We have demonstrated that multiplexing allows to induce stochastic resonance and can be used to control this phenomenon by enhancing or suppressing it. Therefore, multiplexing can play both constructive or destructive role for stochastic resonance. In particular, we show that one can control the regularity of the collective response of coupled oscillators in the regime of stochastic resonance by varying inter-layer coupling strength. In more detail, the stochastic resonance can be suppressed by multiplexing, if not all layers are driven by a common periodic influence. In contrast, multiplexing-based enhancement of stochastic resonance can be achieved when the external force is applied to all the interacting layers. Moreover, the multiplexing-based enhancement of stochastic resonance can be strengthened by increasing the number of interacting layers.

The revealed multiplexing-based enhancement and suppression of stochastic resonance is observed for inter-layer coupling strength being smaller than the intra-layer one. Moreover, the same phenomena are observed when the difference between intra- and inter-layer coupling strength is larger in comparison with the values presented in the paper (results not shown). As demonstrated in Ref. \cite{semenova2018}, weak multiplexing can enhance the effect of coherence resonance. Taking this into account we make here a general conclusion that weak multiplexing can have a sufficiently strong impact on the noise-induced resonant phenomena in multilayer networks and enhance them.

Current research is the first step towards a detailed study of the multiplexing-based stochastic resonance control and raises a number of questions. In particular, the factors determining resonant character of the stochastic resonance enhancement remain to be understood. In addition, by using random initial conditions from the same basin of attraction, we excluded the impact of multistability. A manifold of coexisting stochastic regimes in non-autonomous multilayer networks of bistable oscillators represents another challenging problem. These and other questions are issues for further investigations.

\section*{DATA AVAILABILITY}
The data that support the findings of this study are available from the corresponding author upon reasonable request.

\section*{Acknowledgements}

We are very grateful to professor Alexander Neiman for helpful discussions.
We acknowledge support by the Deutsche Forschungsgemeinschaft (DFG, German Research Foundation) -- Projektnummer -- 163436311-SFB-910. 
V.V.S. also acknowledges support by the Russian Science Foundation (project No.  22-72-00038). 

\nocite{*}

%

\end{document}